\documentclass[a4paper,dvipdfmx]{article}
\usepackage{INTERSPEECH2019,amsmath}
\usepackage{amssymb}
\usepackage{amsfonts}
\usepackage{subfigure}
\usepackage{graphicx}

\renewcommand{\Vec}[1]{\textrm{\boldmath $#1$}} 

\newcommand{\pt}[1]{\left(#1\right)} 
  
\title{
DNN-based Speaker Embedding Using Subjective Inter-speaker Similarity \\
for Multi-speaker Modeling in Speech Synthesis
}
\name{Yuki Saito, Shinnosuke Takamichi, and Hiroshi Saruwatari}
\address{
Graduate School of Information Science and Technology,
The University of Tokyo, \\
7-3-1 Hongo, Bunkyo-ku, Tokyo 113-8656, Japan
}

\email{ \{yuuki\_saito, shinnosuke\_takamichi, hiroshi\_saruwatari\}@ipc.i.u-tokyo.ac.jp}

\begin{document}
\ninept
\setlength{\abovedisplayskip}{8pt}
\setlength{\belowdisplayskip}{8pt}
\setlength\floatsep{5pt}
\setlength\intextsep{5pt}
\setlength\textfloatsep{5pt}

\maketitle
\begin{abstract}
This paper proposes novel algorithms for speaker embedding using subjective inter-speaker similarity based on deep neural networks (DNNs). Although conventional DNN-based speaker embedding such as a $d$-vector can be applied to multi-speaker modeling in speech synthesis, it does not correlate with the subjective inter-speaker similarity and is not necessarily appropriate speaker representation for open speakers whose speech utterances are not included in the training data.  We propose two training algorithms for DNN-based speaker embedding model using an inter-speaker similarity matrix obtained by large-scale subjective scoring. One is based on similarity vector embedding and trains the model to predict a vector of the similarity matrix as speaker representation. The other is based on similarity matrix embedding and trains the model to minimize the squared Frobenius norm between the similarity matrix and the Gram matrix of $d$-vectors, i.e., the inter-speaker similarity derived from the $d$-vectors. We crowdsourced the inter-speaker similarity scores of 153 Japanese female speakers, and the experimental results demonstrate that our algorithms learn speaker embedding that is highly correlated with the subjective similarity. We also apply the proposed speaker embedding to multi-speaker modeling in DNN-based speech synthesis and reveal that the proposed similarity vector embedding improves synthetic speech quality for open speakers whose speech utterances are unseen during the training.
\end{abstract}


\section{Introduction}
Statistical parametric speech synthesis (SPSS)~\cite{zen09} is a technique for synthesizing naturally sounding and easily controllable synthetic speech. Recent developments of both training algorithms and acoustic modeling for SPSS using deep neural networks (DNNs)~\cite{zen13dnn} have significantly improved quality of synthetic speech. For instance, training algorithms based on generative adversarial networks~\cite{goodfellow14} have significantly improved synthetic speech quality by reducing the statistical differences between natural and generated speech parameters~\cite{saito18taslp,saito18icassp-tts,zhao18wgan-tts}. Acoustic model based on Tacotron~\cite{wang17tacotron} and WaveNet vocoder~\cite{oord16wavenet,tamamori17WN-vocoder} has achieved high fidelity natural speech~\cite{shen18}. Although improving the synthetic speech quality is one of the goals of SPSS research, learning interpretable representation for controlling characteristics of the synthetic speech (i.e., features that are highly correlated with human speech perception) is also important. We focus on learning the speaker representation for multi-speaker modeling in DNN-based SPSS.

The simplest way for representing speaker identity in DNN-based SPSS is to use a speaker code~\cite{hojo18spk_code} that denotes one speaker by using a one-hot coded vector. Although the speaker code works well for reproducing the characteristics of \textit{closed} speakers whose speech utterances are included in the training corpora, they have difficulties in 1) dealing with \textit{open} speakers, i.e., previously unseen speakers during the training, because their speaker codes are not defined, and 2) finding a desired speaker when the number of the closed speakers is large. Some techniques for adapting the speaker codes have been proposed for alleviating the first problem~\cite{hojo18spk_code,luong17}; however, the characteristics of the open speakers are not fully reproduced. A more effective approach is utilizing DNNs trained to predict speaker identity from given acoustic features. A $d$-vector~\cite{variani14dvecs} is a well-known example of the DNN-based speaker embedding technique and has been applied to multi-speaker modeling in SPSS~\cite{saito18icassp-vc} using variational auto-encoders (VAEs)~\cite{kingma13vae}. However, even the $d$-vector cannot solve the second problem because it is merely used for verifying a specific speaker, and its coordinates in the embedding space do not correlate with the subjective inter-speaker similarity, i.e., perceptually similar speakers are not necessarily embedded close to each other.

To learn interpretable speaker embedding, we propose novel algorithms incorporating the subjective inter-speaker similarity into training the DNN-based speaker embedding model. First, we conduct large-scale subjective scoring to obtain a matrix representing the subjective inter-speaker similarity. The model is trained to minimize the loss functions defined by the similarity matrix. We investigate two approaches for the training. The first is similarity {\it vector} embedding, which trains the model to predict a vector of the similarity matrix instead of the conventional speaker code. The second is similarity {\it matrix} embedding, which trains the model to minimize the squared Frobenius norm between the similarity matrix and Gram matrix of the $d$-vectors, i.e., inter-speaker similarity derived from the $d$-vectors. We crowdsourced the inter-speaker similarity scores of 153 Japanese female speakers, and the experimental results demonstrate that the proposed algorithms can learn speaker embedding that is highly correlated with the subjective similarity compared with the conventional $d$-vectors. We also investigate the effectiveness of the proposed speaker embedding for the VAE-based multi-speaker SPSS~\cite{saito18icassp-vc}, and reveal that the similarity vector embedding improves synthetic speech quality for open speakers.

\section{Conventional speaker embedding}
\subsection{One-hot speaker code}
A speaker code~\cite{hojo18spk_code}
$\Vec{c} = [c(1), \cdots, c(n), \cdots, c(N_{\rm s})]^{\top}$
is the 1-of-$N_{\rm s}$ representation for identifying the one of closed $N_{\rm s}$ speakers, which is the \textit{discrete} representation of speaker identity. The speaker code 
$\Vec{c}_{i}$
for the $i$th speaker is defined as follows:
\begin{align}
  c_{i}(n) &=
  \begin{cases}
         1 \;\; {\rm if} \;\; n = i \\
         0  \; \; {\rm otherwise} 
  \end{cases} \; (1 \le n \le N_{\rm s}).
\end{align}
Although the one-hot coded representation works reasonably well, it cannot define the identity of open speakers, and the size of the code increases in proportion to the number of the closed speakers. Moreover, it should be difficult for users to find their desired speaker in synthesizing speech because speaker codes completely ignore the subjective inter-speaker similarity.

\subsection{$d$-vector}\label{subsect:dvec}
A $d$-vector~\cite{variani14dvecs} is a bottleneck feature vector extracted from a pre-trained DNN-based speaker recognition model, which is the \textit{continuous} representation of speaker identity. The model is trained to predict speaker identity from a given acoustic feature sequence by minimizing the softmax cross-entropy defined as follows:
\begin{align}
L_{\mathrm{SCE}}\pt{\Vec{c}, \Vec{\hat c}}
& = - \sum_{n = 1}^{N_{\rm s}} c\pt{n} \log {\hat c}\pt{n} \label{eq:L_SCE},
\end{align}
where
$\Vec{\hat c} = [\hat{c}(1), \cdots, \hat{c}(n), \cdots, \hat{c}(N_{\rm s})]^{\top}$
is an output vector of the DNNs.
The $N_{\rm d}$-dimensional $d$-vector
$\Vec{d} = [d(1), \cdots, d(N_{\rm d})]^{\top}$
is extracted from a bottleneck layer of the DNNs. The one before the output layer is often used. Typically, $N_{\rm d}$ is smaller than $N_{\rm s}$ and the $d$-vector enables us to use the lower-dimensional speaker representation. We can define the identity of the $i$th speaker $\Vec{d}_{i}$ as a $d$-vector averaged over all $d$-vectors generated from acoustic features of the $i$th speaker. Although we can embed speakers in the continuous space defined by $d$-vectors and can deal with open speakers~\cite{saito18icassp-vc}, it is still difficult for users to interpret what the speaker embedding means because its coordinates do not correlate with the subjective inter-speaker similarity.

\section{Proposed speaker embedding}
Here, we propose two algorithms for learning speaker embedding that is highly correlated with the subjective inter-speaker similarity.
\subsection{Subjective inter-speaker similarity matrix}
\begin{figure}[t]
  \centering
  \includegraphics[width=0.98\linewidth]{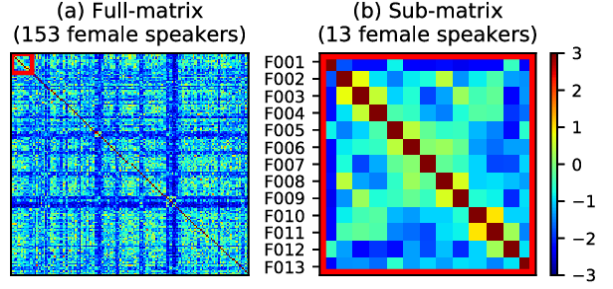}
  \vspace{-8pt}
  \caption{(a) Similarity matrix of 153 Japanese female speakers and (b) its sub-matrix obtained by large-scale subjective scoring.}
  \label{fig:simmat}
\end{figure}

We define a subjective inter-speaker similarity matrix that represents the speaker-pair similarity perceived by listeners.
Let
$\Vec{\rm S} = [\Vec{s}_{1}, \cdots, \Vec{s}_{i}, \cdots, \Vec{s}_{N_{\rm s}}]$
be an
$N_{\rm s}$-by-$N_{\rm s}$
symmetric similarity matrix
and
$\Vec{s}_{i} = [s_{i, 1}, \cdots, s_{i, j}, \cdots, s_{i, N_{\rm s}}]^{\top}$
be an $N_{\rm s}$-dimensional similarity vector of the $i$th speaker. Each element $s_{i,j}$ takes a value between $-v$ and $v$, which represents the perceptual similarity of the $i$th and $j$th speakers. Namely, $s_{i, j}$ stores the average score of the subjective evaluation asking ``To what degree do the $i$th speaker's voice and the $j$th speaker's one sound similar?'' We assume that the diagonal elements, i.e., intra-speaker similarity, take the maximum value of the similarity. Figures~\ref{fig:simmat}(a) and (b) show the similarity matrix of 153 Japanese female speakers and its sub-matrix, respectively. Section 4.1.1 describes details of the subjective scoring for obtaining the score matrix, and Section 4.2 presents analysis of the scores.

\begin{figure}[t]
  \centering
  \includegraphics[width=0.98\linewidth]{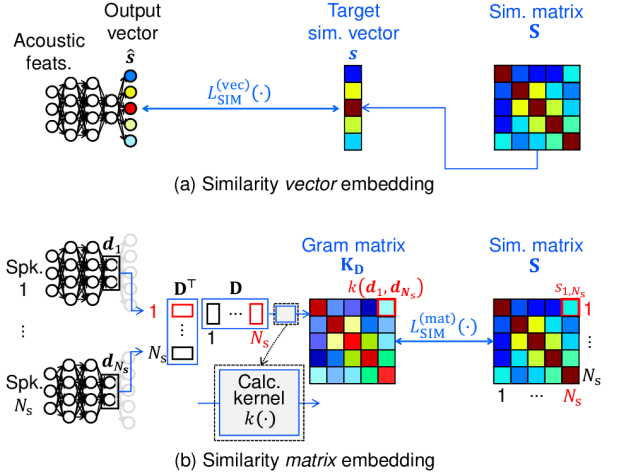}
  \vspace{-8pt}
  \caption{Calculation of loss functions in proposed algorithms based on (a) similarity vector embedding and (b) similarity matrix embedding.}
  \label{fig:prop_vecmat}
\end{figure}

\subsection{Training based on similarity vector embedding}\label{subsect:prop_vec}
The first proposed algorithm uses the similarity vector as a target to be predicted by the DNNs, instead of the conventional speaker code. The loss function for the training is defined as follows:
\begin{align}
L_{\mathrm{SIM}}^{\mathrm{(vec)}}\pt{\Vec{s}, \Vec{\hat s}}
& = \frac{1}{N_{\rm s}} \pt{\Vec{\hat s} - \Vec{s} }^{\top} \pt{ \Vec{\hat s} - \Vec{s} } \label{eq:L_sim_vec},
\end{align}
where $\Vec{s} \in \Vec{\rm S}$ and $\Vec{\hat s}$ denote the target similarity vector and output vector of the DNNs, respectively. Figure~\ref{fig:prop_vecmat}(a) shows the computation procedure of $L_{\mathrm{SIM}}^{\mathrm{(vec)}}(\Vec{\cdot})$.

\subsection{Training based on similarity matrix embedding}\label{subsect:prop_mat}
The second proposed algorithm directly uses the similarity matrix $\Vec{\rm S}$ as a constraint on coordinates of $d$-vectors.
Let
$\Vec{\rm D} = [\Vec{d}_{1}, \cdots, \Vec{d}_{i}, \cdots, \Vec{d}_{N_{\rm s}}]$
be an
$N_{\rm s}$-by-$N_{\rm s}$
matrix including $d$-vectors extracted from all closed speakers.
The loss function for the training is defined as follows:
\begin{align}
L_{\mathrm{SIM}}^{\mathrm{(mat)}} \pt{ \Vec{\rm D}, \Vec{\rm S} }
&= \frac{2}{ \left\| \Vec{1}_{N_{\rm s}} - \Vec{\rm I}_{N_{\rm s}} \right\|_{F}^{2} } \left\| \widetilde{ \Vec{\rm K} }_{ \Vec{\rm D} } - \widetilde{ \Vec{\rm S} } \right\|_{F}^{2}
\label{eq:L_sim_mat}, \\
\widetilde{\Vec{\rm K}}_{\Vec{\rm D}} &= \Vec{\rm K}_{ \Vec{\rm D} } - \pt{ \Vec{\rm K}_{ \Vec{\rm D} } \odot \Vec{\rm I}_{N_{\rm s}} }, \\
\widetilde{\Vec{\rm S}} &= \Vec{\rm S} - s \Vec{\rm I}_{N_{\rm s}},
\end{align}
where
$\| \Vec{\cdot} \|_{F}^{2}$,
$\odot$,
$\Vec{1}_{N_{\rm s}}$,
and
$ \Vec{\rm I}_{N_{\rm s}} $
denote
the squared Frobenius norm of the given matrix,
Hadamard product,
$N_{\rm s}$-by-$N_{\rm s}$
matrix whose all components are 1,
and
$N_{\rm s}$-by-$N_{\rm s}$
identity matrix, respectively.
$2 / \| \Vec{1}_{N_{\rm s}} - \Vec{\rm I}_{N_{\rm s}} \|_{F}^{2}$
is a normalization coefficient corresponding to the degrees of freedom of the matrix $\widetilde{\Vec{\rm K}}_{\Vec{\rm D}} - \widetilde{\Vec{\rm S}}$.
$\Vec{\rm K}_{\Vec{\rm D}}$ is the Gram matrix of $d$-vectors defined as:
\begin{align}
  \Vec{\rm K}_{\Vec{\rm D}} = \left[
    \begin{array}{ccc}
      k\pt{ \Vec{d}_{1}, \Vec{d}_{1} } & \cdots & k\pt{ \Vec{d}_{1}, \Vec{d}_{N_{\rm s}} }  \\
      \vdots & \ddots & \vdots \\
      k\pt{ \Vec{d}_{N_{\rm s}}, \Vec{d}_{1} } & \cdots & k\pt{ \Vec{d}_{N_{\rm s}} \Vec{d}_{N_{\rm s}}}
    \end{array}
  \right],
\end{align}
where $k(\Vec{d}_{i}, \Vec{d}_{j})$ is a kernel function calculated by using $\Vec{d}_{i}$ and $\Vec{d}_{j}$, i.e., the speaker similarity derived from the $d$-vectors. This proposed algorithm can directly learn speaker embeddings correlated with the subjective inter-speaker similarity. Figure~\ref{fig:prop_vecmat}(b) shows the computation procedure of $L_{\mathrm{SIM}}^{\mathrm{(mat)}}(\Vec{\cdot})$.

The minimization of Eq.~(\ref{eq:L_sim_mat}) means the speaker pairs are embedded with the consideration of both their perceptual similarity and dissimilarity. For the actual use of controllable DNN-based SPSS, at least the consideration of the similar speaker pairs should be satisfied. Therefore, we can relax Eq.~(\ref{eq:L_sim_mat}) to satisfy this as follows:
\begin{align}
L_{\mathrm{SIM}}^{\mathrm{(mat-re)}} \pt{ \Vec{\rm D}, \Vec{\rm S} }
&= \frac{2}{ \left\| \Vec{\rm W} - \Vec{\rm I}_{\rm s} \right\|_{F}^{2} } \left\| \Vec{\rm W} \odot \pt{ \widetilde{ \Vec{\rm K} }_{ \Vec{\rm D} } - \widetilde{ \Vec{\rm S} } } \right\|_{F}^{2}
\label{eq:L_sim_mat_re},
\end{align}
where $\Vec{\rm W} = [w_{i, j}]_{1 \le i, j \le N_{\rm s}}$ is defined as $w_{i, j} = 1$ if $s_{i, j} > 0$ otherwise 0 and $2/\| \Vec{\rm W} - \Vec{\rm I}_{\rm s} \|_{F}^{2}$ is a normalization coefficient corresponding to the degrees of freedom of the matrix
$\Vec{\rm W}$.
In this formulation, perceptually dissimilar speaker pairs are not considered in training.

\begin{figure}[t]
  \centering
  \includegraphics[width=0.98\linewidth]{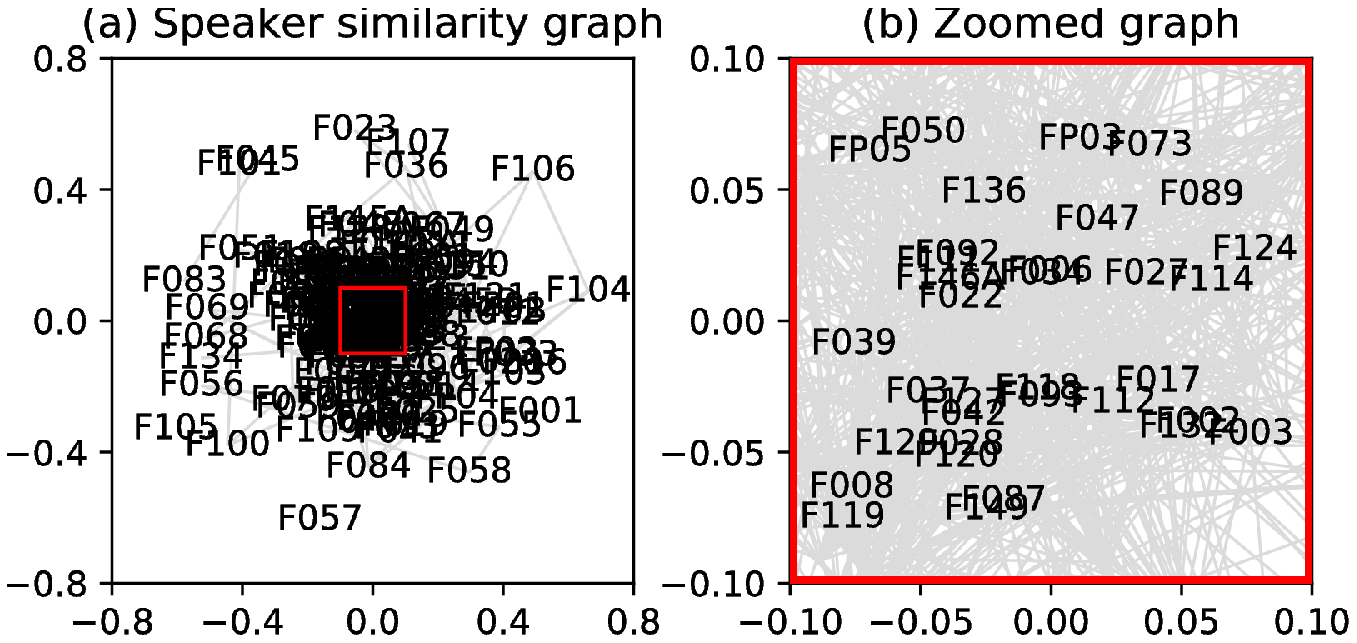}
  \vspace{-8pt}
  \caption{(a) Speaker similarity graph defined by similarity matrix shown in Fig.~\ref{fig:simmat}(a) and (b) its zoomed version.}
  \label{fig:graph}
\end{figure}

\subsection{Discussion}
The similarity matrix used in the proposed algorithms offers better understanding of the relationships among speakers by visualizing them as a graph defined by the matrix. Figures~\ref{fig:graph}(a) and (b) show the speaker similarity graph defined by the similarity matrix shown in Fig.~\ref{fig:simmat}(a) and its zoomed version, respectively. The adjacency matrix of the graph was the same as the matrix $\Vec{\rm W}$ of Eq.~(\ref{eq:L_sim_mat_re}). The positions of the speakers in Fig.~\ref{fig:graph} were determined by using multidimensional scaling with the similarity matrix. The minimization of Eq.~(\ref{eq:L_sim_mat_re}) is similar to learning the graph from acoustic features. Therefore, we expect to further introduce graph signal processing~\cite{shuman13} and graph embedding~\cite{goyal18} to DNN-based speaker embedding and multi-speaker modeling in SPSS.

Regarding prior works, Tachibana et al.~\cite{tachibana06} and Ohta et al.~\cite{ohta07} proposed controllable SPSS in the hidden Markov model (HMM) and Gaussian mixture model (GMM) era. They modeled the characteristics of a specific speaker with subjective impressions such as "warm -- cold" and "clear -- hoarse" so that the latent variables of the HMMs or GMMs were related to the word pairs. Our algorithms extend these ideas to make the DNNs model the {\it pair-wise} speaker similarity as the embedding vectors rather than the conventional {\it point-wise} impressions of one speaker. Furthermore, the relationship between the speakers' intention and the listeners' perception (e.g., the differences in emotion perception~\cite{lorenzo-trueba18}) can be modeled by our algorithms.

The proposed algorithm based on similarity matrix embedding can directly learn inter-speaker relationships by making the Gram matrix of $d$-vectors close to the similarity matrix. We can choose an arbitrary kernel function to construct the embedding space. When the inner product is used as the kernel function, Eq.~(\ref{eq:L_sim_mat}) is equivalent to deep clustering~\cite{hershey16} (except for subtracting the diagonal components). Not only such a simple kernel but also a more complicated one can be utilized in our method.

\section{Experimental Evaluation}
\subsection{Experimental conditions}\label{subsect:exp_cond}
\subsubsection{Conditions for large-scale subjective scoring}\label{subsubsect:subjective_eval}
We conducted large-scale subjective scoring for obtaining the similarity matrix $\Vec{\rm S}$ by using our crowdsourced evaluation systems. We used 153 Japanese female speakers included in the JNAS corpus~\cite{jnas}. Each speaker utters at least 150 reading-style utterances (totally about 44 hours). We extracted five non-parallel utterances per speaker for scoring the text-independent inter-speaker similarity of the 153 speakers. Each listener scored the similarity of 34 randomly selected speaker pairs extracted from all of the possible 11,628 different speaker pairs. The score was an integer between $-3$ (completely different) and $+3$ (very similar). The similarity of one of the 11,628 speaker pairs was scored by at least 10 different listeners. Finally, 4,060 listeners participated in the scoring, and 138,040 answers were obtained.

\begin{figure}[t]
  \centering
  \includegraphics[width=0.98\linewidth]{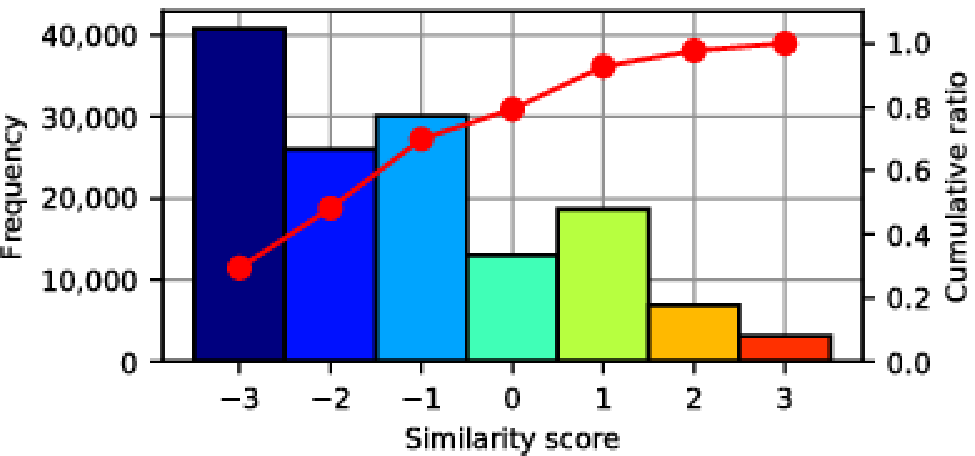}
  \vspace{-8pt}
  \caption{Histogram of similarity scores with its cumulative ratio denoted by red line.}
  \label{fig:hist}
\end{figure}

\begin{figure*}[t]
  \centering
  \includegraphics[width=0.98\linewidth]{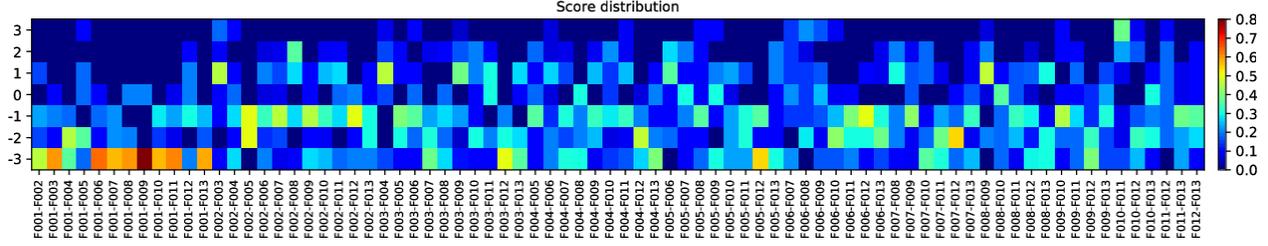}
  \vspace{-8pt}
  \caption{Histogram of speaker-pair-wise similarity scores. The speaker pair includes the 13 speakers shown in  Fig.~\ref{fig:simmat}(b).}
  \label{fig:dist}
\end{figure*}

\subsubsection{Conditions for DNN-based speaker embedding}\label{subsubsect:cond_embed}
The JNAS corpus was also used for training DNN-based speaker embedding model. Ninety percent of the utterances and the remainder were used for training and evaluation, respectively. The five utterances used for the subjective scoring were omitted from both the training and evaluation data. The number of utterances per speaker was balanced among the speakers. The number of closed speakers used for training, $N_{\rm s}$, was set to 140, except for 13 speakers shown in Fig.~\ref{fig:simmat}(b). The 13 open speakers were used for objective and subjective evaluations in Sections \ref{subsect:obj} and \ref{subsect:sbj_spss}. Although each element in the similarity matrix was ranged in [$-3$, $+3$], we normalized the values to be in [$-1$, $+1$] during the training. Accordingly, the sigmoid kernel $k(\Vec{d}_{i}, \Vec{d}_j) = \mathrm{tanh}( \Vec{d}_{i}^{\top} \Vec{d}_{j})$ was used for the proposed similarity matrix embedding described in Section~\ref{subsect:prop_mat}. The $d$-vector of a specific speaker was estimated as the average value of $d$-vectors extracted from acoustic features in the voiced region. The voiced/unvoiced decision was obtained from an $F_0$ sequence extracted by STRAIGHT vocoder systems~\cite{kawahara99}.

The DNN architecture for speaker embedding model was Feed-Forward networks that included 4 hidden layers with the tanh activation function. The number of hidden units at the 1st-through-3rd layers and the 4th layer for extracting $d$-vectors were 256 and 8, respectively. The softmax activation function was used for the output layer of the embedding model in the algorithms described in Sections \ref{subsect:dvec} and \ref{subsect:prop_mat}. Since the values of the similarity matrix $\Vec{\rm S}$ were normalized in [$-1$, $+1$], the tanh activation function was applied to the output layer of the embedding model in the proposed similarity vector embedding described in \ref{subsect:prop_vec}. The input of the embedding model was the joint vectors of the 1st-through-39th mel-cepstral coefficients and their dynamic features, and they were normalized to have zero-mean unit-variance during the training. The mel-cepstral coefficients were extracted by using STRAIGHT vocoder systems~\cite{kawahara99}. AdaGrad~\cite{adagrad} was used as the optimization algorithm, setting its learning rate to 0.01. All training algorithms described in Sections \ref{subsect:dvec}, \ref{subsect:prop_vec}, and \ref{subsect:prop_mat} were performed with 100 epochs.

\subsubsection{Conditions for VAE-based SPSS}\label{subsubsect:cond_spss}
We constructed the VAE-based SPSS~\cite{saito18icassp-vc} that incorporated DNN-based speech recognition and speaker embedding models into speech synthesis for achieving high-quality multi-speaker modeling. The DNN architecture for the recognition model was Feed-Forward networks that included 4 hidden layers with the tanh activation function. The number of hidden units was 1,024. The recognition model was trained to output 43-dimensional Japanese phonetic posteriorgrams (PPGs)~\cite{sun16} from the same input vector as the speaker embedding model, i.e., the 78-dimensional acoustic feature vector including static-dynamic mel-cepstral coefficients. The recognition model training was performed with 100 epochs. About 50 utterances per one in the 140 closed speakers were used for the recognition model training. The embedding model was the same as the DNNs described in Section \ref{subsubsect:cond_embed}. The DNN architecture for the VAEs was Feed-Forward networks that consisted of encoder and decoder networks. The encoder had two hidden layers with the rectified linear unit (ReLU)~\cite{glorot11relu} activation function and extracted the 64-dimensional latent variables from a joint vector of the static-dynamic mel-cepstral coefficients and PPGs. The first and second hidden layers had 256 and 128 hidden units, respectively. The decoder reconstructed the input static-dynamic mel-cepstral coefficients from a joint vector of the latent variables, PPGs, and the 8-dimensional speaker embedding vector. The DNN architecture for the decoder was symmetric about that for the encoder. The VAEs were trained to maximize the variational lower bound of the log likelihood~\cite{kingma13vae} with 25 epochs using the same training data as in the embedding model training. The maximum likelihood parameter generation~\cite{tokuda00} was performed to generate static mel-cepstral coefficients considering their temporal dependencies. The generated mel-cepstral coefficients, input F0, and 5 band-aperiodicity~\cite{kawahara01,ohtani06} were used for synthesizing speech waveform based on the STRAIGHT vocoder systems~\cite{kawahara99}.

\begin{figure}[t]
  \centering
  \includegraphics[width=0.99\linewidth]{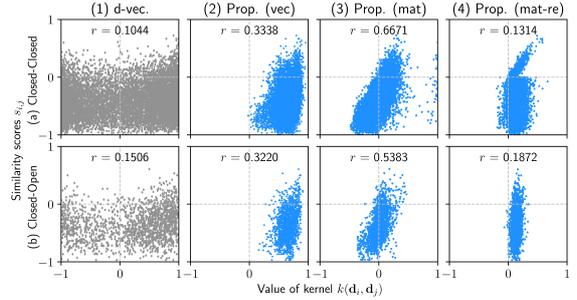}
  \vspace{-8pt}
  \caption{Scatter plots of similarity scores $s_{i, j}$ and values of kernel $k(\Vec{d}_{i}, \Vec{d}_{j})$ with their correlation coefficient $r$. These plots were made by all speaker pairs.}
  \label{fig:cor}
\end{figure}

\begin{figure}[t]
  \centering
  \includegraphics[width=0.99\linewidth]{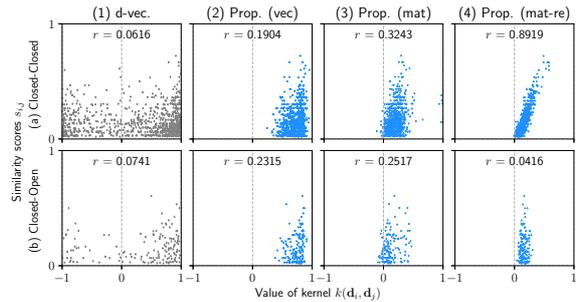}
  \vspace{-8pt}
  \caption{Scatter plots of similarity scores $s_{i, j}$ and values of kernel $k(\Vec{d}_{i}, \Vec{d}_{j})$ with their correlation coefficient $r$. These plots were made by speaker pairs whose similarity scores were greater than 0.}
  \label{fig:cor_only_sim}
\end{figure}

\subsection{Analysis of crowdsourced similarity scores}
We analyzed the crowdsourced similarity scores that made the similarity matrix shown in Fig.~\ref{fig:simmat}(a). The histogram of the all crowdsourced scores is shown in Fig.~\ref{fig:hist}. From this figure, we found that about 70\% of the scores were smaller than zero. Also, Fig.~\ref{fig:dist} plots a histogram of speaker-pair-wise scores. From this figure, we observed that most of the listeners scored $-3$ for more dissimilar speaker pairs (e.g., ``F001-F009''). On the other hand, listeners scored more various values for more similar speaker pairs (e.g., ``F010-F011''). These results suggested that listeners could easily find the dissimilar speakers rather than similar ones.

\subsection{Objective evaluation of speaker embedding}~\label{subsect:obj}
To investigate the correlation between the subjective similarity score and speaker embedding, we computed the Pearson correlation coefficient between the similarity scores $s_{i, j}$ and the values of the kernel function $k(\Vec{d}_{i}, \Vec{d}_j)$. We compared the following four algorithms:
\begin{description}
\item[d-vec.]: Eq.~(\ref{eq:L_SCE})
\item[Prop. (vec)]: Eq.~(\ref{eq:L_sim_vec})
\item[Prop. (mat)]: Eq.~(\ref{eq:L_sim_mat})
\item[Prop. (mat-re)]: Eq.~(\ref{eq:L_sim_mat_re})
\end{description}

The results with their scatter plots are shown in Fig.~\ref{fig:cor}. We found that the conventional $d$-vectors had a weak correlation with the similarity scores. Meanwhile, the three proposed algorithms trained the speaker embedding models so that the embedding vectors had strong correlations with the inter-speaker similarity, which demonstrated that the algorithms could learn speaker embedding that is highly correlated with the subjective scores compared with the conventional $d$-vectors. Focusing on the speaker pairs whose similarity scores were greater than 0 (shown in Fig.~\ref{fig:cor_only_sim}), ``Prop. (mat-re)'' scored the strongest correlations between the similarity scores and speaker embeddings of the closed speakers among the four algorithms. However, it did not work well in the case of the ``Closed-Open'' speaker pairs. One of the causes might be the data sparsity problem, since the number of the similar speaker pairs was significantly smaller than that of the dissimilar speaker pairs, as shown in Fig.~\ref{fig:hist}.

\begin{table}[t]
\label{tab:eval_dv_nat}
\centering
\caption{Preference scores on naturalness (left: conventional $d$-vector, right: proposed methods)}
\vspace{-8pt}
\footnotesize
\begin{tabular}{|c|c|c|c|}
\hline \rule[0mm]{0mm}{2mm}
Speaker & Prop. (vec) & Prop. (mat) & Prop. (mat-re) \\ \hline \rule[0mm]{0mm}{2mm}
F001 & 0.408 - \textbf{0.592} & 0.456 - \textbf{0.544} & 0.448 - \textbf{0.552} \\ \rule[ 0mm]{0mm}{2mm}
F002 & 0.456 - \textbf{0.544} & 0.456 - \textbf{0.544} & 0.504 - 0.496 \\ \rule[0mm]{0mm}{2mm}
F003 & 0.416 - \textbf{0.584} & 0.444 - \textbf{0.556} & 0.448 - \textbf{0.552} \\ \rule[0mm]{0mm}{2mm}
F004 & 0.452 - \textbf{0.548} & 0.460 - 0.540 & 0.432 - \textbf{0.568} \\ \rule[0mm]{0mm}{2mm}
F005 & 0.380 - \textbf{0.620} & 0.484 - 0.516 & 0.484 - 0.516 \\ \rule[0mm]{0mm}{2mm}
F006 & 0.400 - \textbf{0.600} & 0.452 - \textbf{0.548} & 0.424 - \textbf{0.576} \\ \rule[0mm]{0mm}{2mm}
F007 & 0.424 - \textbf{0.576} & 0.484 - 0.516 & 0.492 - 0.508 \\ \rule[0mm]{0mm}{2mm}
F008 & 0.436 - \textbf{0.564} & 0.384 - \textbf{0.616} & 0.436 - \textbf{0.564} \\ \rule[0mm]{0mm}{2mm}
F009 & 0.428 - \textbf{0.572} & 0.492 - 0.508 & 0.460 - 0.540 \\ \rule[0mm]{0mm}{2mm}
F010 & 0.436 - \textbf{0.564} & 0.464 - 0.536 & 0.452 - \textbf{0.548} \\ \rule[0mm]{0mm}{2mm}
F011 & 0.460 - 0.540 & 0.428 - \textbf{0.572} & 0.452 - \textbf{0.548} \\ \rule[0mm]{0mm}{2mm}
F012 & 0.436 - \textbf{0.564} & 0.460 - 0.540 & 0.524 - 0.476 \\ \rule[0mm]{0mm}{2mm}
F013 & 0.428 - \textbf{0.572} & 0.436 - \textbf{0.564} & 0.412 - \textbf{0.588} \\ \hline \rule[0mm]{0mm}{2mm}
Avg. & 0.428 - \textbf{0.572} & 0.454 - \textbf{0.546} & 0.459 - 0.541 \\
\hline
\end{tabular}
\end{table}

\subsection{Subjective evaluation of VAE-based SPSS}~\label{subsect:sbj_spss}
To investigate the effectiveness of the proposed speaker embedding in the VAE-based SPSS, we conducted subjective evaluations on the naturalness and speaker similarity of the synthetic speech of the 13 open speakers. We generated speech samples using mel-cepstral coefficients predicted by the VAEs trained with the 4 different speaker embedding models. Fifty utterances of the speakers were used for estimating the speaker embedding fed into the decoder of the VAE and evaluating the synthetic speech quality. We conducted a series of preference tests (AB tests) on the naturalness of the synthetic speech that compared the conventional algorithm with the three proposed algorithms. Twenty-five listeners participated in each of the following evaluations by using our crowd-sourced evaluation systems, and each listener evaluated 10 speech samples randomly extracted from the 50 utterances. Similarly, we conducted a series of XAB tests on the speaker similarity of the synthetic speech using the natural speech of the speaker as the reference speech samples. The total number of task sets was 2 (AB or XAB) $\times$ 3 (proposed embedding algorithms) $\times$ 13 (open speakers) $\times$ 25 (listeners per one task) $=$ 1,950.

The preference scores on the naturalness and speaker similarity are shown in Tables 1 and 2, respectively. The bold values denote that there is a significant difference between the two scores ($p < 0.05$). The row ``Avg.'' means the scores averaged over all speakers. From the results, we found that ``Prop. (vec)'' always improved both the naturalness and speaker similarity of the synthetic speech, which indicated that the proposed speaker embedding considering the subjective inter-speaker similarity was effective for multi-speaker modeling in DNN-based SPSS. We observed that ``Prop. (mat)'' also improved the naturalness; however, it significantly degraded the speaker similarity in some cases (e.g., ``F005'' and ``F012''). Similar tendencies were observed in the scores of ``Prop. (mat-re).'' To investigate the reason, we calculated the degree of a vertex of the speaker similarity graph shown in Fig.~\ref{fig:graph}(a), i.e., the number of similar speakers of each speaker, and found that those of ``F005'' and ``F012'' were 7 and 1, respectively. Therefore, we inferred that the proposed similarity matrix embedding might not work well when the number of similar speakers was small. Also, the lower generalization towards the open speakers of ``Prop. (mat-re)'' might cause the degradation of the speaker similarity, as shown in Fig.~\ref{fig:cor_only_sim}(b)(4).

\begin{table}[t]
\label{tab:eval_dv_sim}
\centering
\caption{Preference scores on speaker similarity (left: conventional $d$-vector, right: proposed methods)}
\vspace{-8pt}
\footnotesize
\begin{tabular}{|c|c|c|c|}
\hline \rule[0mm]{0mm}{2mm}
Speaker & Prop. (vec) & Prop. (mat) & Prop. (mat-re) \\ \hline \rule[0mm]{0mm}{2mm}
F001 & 0.436 - \textbf{0.564} & 0.488 - 0.512 & 0.528 - 0.472 \\ \rule[ 0mm]{0mm}{2mm}
F002 & 0.468 - 0.532 & 0.496 - 0.504 & 0.488 - 0.512 \\ \rule[0mm]{0mm}{2mm}
F003 & 0.432 - \textbf{0.568} & 0.504 - 0.496 & \textbf{0.604} - 0.396 \\ \rule[0mm]{0mm}{2mm}
F004 & 0.380 - \textbf{0.620} & 0.404 - \textbf{0.596} & 0.488 - 0.512 \\ \rule[0mm]{0mm}{2mm}
F005 & 0.428 - \textbf{0.572} & \textbf{0.616} - 0.384 & \textbf{0.596} - 0.404 \\ \rule[0mm]{0mm}{2mm}
F006 & 0.428 - \textbf{0.572} & 0.444 - \textbf{0.556} & 0.464 - 0.536 \\ \rule[0mm]{0mm}{2mm}
F007 & 0.492 - 0.508 & \textbf{0.568} - 0.432 & \textbf{0.548} - 0.452 \\ \rule[0mm]{0mm}{2mm}
F008 & 0.424 - \textbf{0.576} & 0.500 - 0.500 & 0.504 - 0.496 \\ \rule[0mm]{0mm}{2mm}
F009 & 0.400 - \textbf{0.600} & 0.500 - 0.500 & 0.448 - \textbf{0.552} \\ \rule[0mm]{0mm}{2mm}
F010 & 0.432 - \textbf{0.568} & 0.404 - \textbf{0.596} & 0.496 - 0.504 \\ \rule[0mm]{0mm}{2mm}
F011 & 0.348 - \textbf{0.652} & 0.444 - \textbf{0.556} & 0.536 - 0.464 \\ \rule[0mm]{0mm}{2mm}
F012 & 0.492 - 0.508 & \textbf{0.544} - 0.456 & \textbf{0.564} - 0.436 \\ \rule[0mm]{0mm}{2mm}
F013 & 0.372 - \textbf{0.628} & \textbf{0.564} - 0.436 & 0.452 - \textbf{0.548} \\ \hline \rule[0mm]{0mm}{2mm}
Avg. & 0.426 - \textbf{0.574} & 0.498 - 0.502 & 0.517 - 0.483 \\
\hline
\end{tabular}
\end{table}

\section{Conclusion}
This paper proposed novel algorithms for incorporating subjective inter-speaker similarity perceived by listeners into the training a speaker embedding model based on deep neural networks (DNNs). The algorithms used an inter-speaker similarity matrix obtained from large-scale subjective scoring as a constraint on training the model. Two approaches for the training were investigated. One is similarity {\it vector} embedding, which trains the model to predict a vector of the similarity matrix. The other is similarity {\it matrix} embedding, which trains the model to minimize the squared Frobenius norm between the similarity matrix and Gram matrix of speaker embeddings. For obtaining the similarity matrix, we conducted large-scale subjective scoring in terms of inter-speaker similarity. The experimental results of the DNN-based speaker embedding using the scores demonstrated that the proposed algorithms learned speaker embedding that is highly correlated with the subjective similarity compared with the conventional $d$-vectors. We also investigated the effectiveness of our speaker embedding for multi-speaker modeling in DNN-based speech synthesis, and found that the proposed similarity vector embedding improved naturalness and speaker similarity of the synthetic speech. In the future, we will investigate the effects of the kernel function and the parameterization of the similarity score (e.g., using the interval $[0, 1]$, where 1 means "similar" while 0 means "dissimilar") in the proposed similarity matrix embedding, and improve the speaker similarity of the algorithm by introducing more sophisticated techniques of graph signal processing such as graph convolutional networks~\cite{kipf16gcn} to the training.

\section{Acknowledgements}
Part of this work was supported by the SECOM Science and Technology Foundation, JSPS KAKENHI Grant Number 18J22090 and 17H06101, the Ministry of Internal Affairs and Communications, and the GAP foundation program of the University of Tokyo.


%
%
%


\bibliographystyle{IEEEbib}
\bibliography{Template.aux}

\end{document}